\documentclass[12pt]{article}
\def\be{\begin{equation}}
\def\ee{\end{equation}}
\def\bea{\begin{eqnarray}}
\def\eea{\end{eqnarray}}

\begin{document}
\title{Semi-leptonic Octet Baryon Weak Axial-Vector Form Factors in the Chiral Constitutent Quark Model}
\author{Harleen Dahiya$^a$, Neetika Sharma$^a$, P.K. Chatley$^a$ and Manmohan Gupta$^b$    \\
{\small{\it $^a$ Department of Physics, Dr. B.R. Ambedkar National
Institute of Technology,}}\\
 {\small{\it Jalandhar, Punjab-144 011, India.}}\\
{\small {\it $^b$Department of Physics, Centre of Advanced Study
in Physics,}} \\ {\small{\it Panjab University, Chandigarh-160
014, India.}}} \maketitle

\begin{abstract}

The weak vector and axial-vector form factors have been
investigated for the semi-leptonic octet baryon decays in the
chiral constituent quark model with configuration mixing
($\chi$CQM$_{{\rm config}}$). The effects of SU(3) symmetry
breaking and configuration mixing have also been investigated and
the results are not only in good agreement with existing
experimental data but also show improvement over other
phenomenological models.

\end{abstract}

\maketitle

The measurements of the polarized struture functions in the deep
inelastic scattering (DIS) experiments indicate that the valence
quarks of the proton carry only about 30\% of its spin and also
establishes the asymmetry of the quark distribution functions
\cite{emc}. This is referred to as the ``proton spin problem'' and
it relates the DIS measurements to the weak vector and
axial-vector form factors ($f_{i=1,2,3}(Q^2$) and
$g_{i=1,2,3}(Q^2$)) of the semi-leptonic baryon decays
\cite{okun}. The data to study the form factors in the early
eighties was analyzed under the assumptions of exact SU(3)
symmetry \cite{cern-WA2}. However, the experiments performed
recently \cite{syh} are more precise and indicate that SU(3)
symmetry breaking effects are important. The broader question of
SU(3) symmetry breaking  has been recently discussed by several
authors in various phenomenological models \cite{models} however,
the results of all these models are not in agreement with each
other on the magnitude as well as sign of these form factors.
Pending further experiments, it is of interest to estimate the
role of SU(3) symmetry breaking in the weak form factors.

Chiral constituent quark model with configuration
($\chi$CQM$_{{\rm config}}$) \cite{manohar,cheng,hd}  can yield an
adequate description of the ``quark sea'' generation through the
chiral fluctuations and has been successful in explaining various
general features of the quark flavor and spin distribution
functions and baryon magnetic moments. It, therefore, becomes
desirable to carry out a detailed phenomenological analysis within
the framework of $\chi$CQM$_{{\rm config}}$ to calculate the
axial-vector form factors \cite{tommy}. The purpose of the present
communication is to carry out a detailed analysis of the weak
vector and axial-vector form factors at low energies for the
baryon semi-leptonic decays within the framework of
$\chi$CQM$_{{\rm config}}$ and to understand the role of SU(3)
symmetry breaking and constituent quark masses.

In the $\chi$CQM formalism \cite{cheng}, the fluctuation process
is $q^{\pm} \rightarrow {\rm GB}
  + q^{' \mp} \rightarrow  (q \bar q^{'})
  +q^{'\mp}$, where GB represents the Goldstone boson and $q \bar q^{'}  +q^{'}$
 constitute the ``quark sea''.
The effective Lagrangian describing interaction between quarks and
a nonet of GBs can be expressed as ${\cal L}= g_8 { \bf \bar
q}\left(\Phi\right) {\bf q}$, where $g_8$ is the coupling
constants and $\Phi$ is the GB field. The SU(3) symmetry breaking
is introduced by considering $m_s > m_{u,d}$ as well as by
considering the masses of GBs to be nondegenerate $(M_{K,\eta} >
M_{\pi})$. The nucleon wavefunction with configuration mixing can
be expressed as $|B\rangle = \cos \phi |56,0^+\rangle_{N=0} +
\sin\phi|70,0^+\rangle_{N=2}$, where $\phi$ represents the mixing
angle. The spin structure of a nucleon is defined as $\hat B
\equiv \langle B|N|B\rangle$ \cite{{cheng}}, where $N$ is the
number operator $ N=n_{u^{+}}u^{+} + n_{u^{-}}u^{-} +
n_{d^{+}}d^{+} + n_{d^{-}}d^{-} + n_{s^{+}}s^{+} +
n_{s^{-}}s^{-}$, $n_{q^{\pm}}$ being the number of $q^{\pm}$
quarks.

The transition matrix element for the semi-leptonic hadronic decay
process $B_i \rightarrow B_f +l +\bar {\nu_l}$ is given as $M
=\frac{G_F}{\sqrt2} (V_{q_iq_f}) \langle B_f|J{^\mu_h}|B_i\rangle
(\bar u_l(p_l)\gamma_\mu(1- \gamma_5) u_\nu(p_\nu))$. The weak
hadronic current can be expressed in terms of the vector and
axial-vector current as $\langle
B_f(p_{f})|J{^\mu_h}|B_i(p_{i})\rangle= \langle B_f
(p_{f})|J{^\mu_V}-J{^\mu_A} |B_i(p_{i}) \rangle$ and the
respective matrix elements for the vector and axial-vector current
are given as
\be
\bar u_f(p_f) \left( f_1(Q^2) \gamma^\mu -
i\frac{f_2(Q^2)}{M_i+M_f} \sigma^{\mu\nu} q_\nu
+\frac{f_3(Q^2)}{M_i + M_f}q^\nu \right) u_i(p_i)\,, \label{jv}
\ee
\be
 \bar u_f(p_f)\left (
g_1(Q^2)\gamma ^\mu \gamma^5  -i\frac{g_2(Q^2)}{M_i +
M_f}\sigma^{\mu\nu} q_\nu \gamma^5 +\frac{g_3(Q^2)}{M_i +
M_f}q^\nu \gamma^5 \right) u_i(p_i)\,,\label{ja} \ee where  $M_i$
$(M_f)$ and $u_i(p_i)$ ($\bar u_f(p_f)$) are the masses and Dirac
spinors of the initial (final) baryon states, respectively. The
four momenta transfer is given as $Q^2 = -q^2$, where $ q \equiv
p_i - p_f$. The functions $f_i(Q^2)$ and $g_i(Q^2)$ ($i=1,2,3$)
are the dimensionless vector and axial-vector form factors
required to be real by $G-$parity. These equations can be solved
using Gordon equality, in the Breit-frame (Lorentz frame), where
${\bf p_i} = -{\bf p_f}= \frac{1}{2}{\bf q}$ in the
nonrelativistic limit $({\bf q}^2 \ll M_i^2 ,M_f^2 )$ , giving the
vector and axial-vector form factors in terms of the Sachs form
factors at $Q^2\approx 0$.

Further, the linear part of form factors can be solved in the
quark sector and the form factors for the quark currents can be
used to obtain the corresponding vector and axial-vector form
factors for the baryons which are expressed as \be f_1 =
f_1(0)\,,~~~ f_2 = \left(\frac{\Sigma M}{\Sigma m
}\frac{G_A}{G_V}-1\right)f_1(0)\,, ~~~ f_3 = \frac{\Sigma
M}{\Sigma m}\left(E \frac{G_A}{G_V} -\epsilon\right)f_1(0)\,,\ee
\be g_1 = g{_1}(0)\,, ~~g_2 = \left(\frac{\Sigma M}{\Sigma
m}\epsilon -\frac{1}{2}(1+\frac{\Sigma M^2}{\Sigma m^2})E\right)
g_1(0)\,, ~~g_3 = \left( \frac{1}{2}(1-\frac{\Sigma M^2}{\Sigma
m^2} )+\frac{\Sigma M ^2}{\Sigma m^2}g^q_3\right)g_1(0)\,, \ee
where the higher order terms involving $E \equiv \frac{\Delta
M}{\Sigma M}$ and $\epsilon\equiv \frac{\Delta m}{\Sigma m} $ have
been neglected. The ratio of $g_1$ and $f_1$ is the non-singlet
combination of the quark spin polarizations given as
$\Delta_3=\frac{G_A}{G_V} =\frac{g_1(0)}{f_1(0)}$. The hyperon
decays considered in the present work are
 $n\rightarrow p$, $\Sigma^ \mp \rightarrow \Lambda$,
$\Sigma^-\rightarrow \Sigma^0$ and $ \Xi^- \rightarrow \Xi^0$
corresponding to the strangeness conserving decays and $\Sigma^-
\rightarrow n$, $\Xi^- \rightarrow \Sigma^0$, $\Xi^- \rightarrow
\Lambda$, $\Lambda \rightarrow p$ and $\Xi^0\rightarrow \Sigma^+$
corresponding to the strangeness changing decays.

The $\chi$CQM$_{{\rm config}}$ involves five parameters, four of
these $a$, $a \alpha^2$, $a \beta^2$, $a \zeta^2$ representing
respectively the probabilities of fluctuations to pions, $K$,
$\eta$, $\eta^{'}$, following the hierarchy $a > \alpha > \beta >
\zeta$, while the fifth representing the mixing angle. The mixing
angle $\phi$ is fixed from the consideration of neutron charge
radius, whereas the other parameters have been fitted using
$\Delta u$, $\Delta_3$ \cite{PDG} as well as $\bar u-\bar d$,
$\bar u/\bar d$ \cite{emc} leading to $a=0.12$, $\zeta=-0.15$,
$\alpha=\beta=0.45$ as the best fit values.

In Table \ref{fsandgs}, we have given the individual values of
vector and axial-vector form factors in the $\chi$CQM$_{{\rm
config}}$. Some important conclusions can be drawn regarding the
dependence of these form factors on configuration mixing and SU(3)
symmetry breaking. The ratio $\frac{g_1}{f_1}$ ratio depends on
configuration mixing as well as  SU(3) symmetry breaking. The
other form factors $f_2$, $f_3$, $g_2$ and $g_3$ are found to be
dependent on the constituent quark masses only. It can be clearly
seen from the results that the contributions of second class
currents $f_3$ and $g_2$ are very small for the same isospin
multiplets. This is because of the small mass difference between
the initial and final decay particles.

In Table \ref{g1f1}, we have presented the values of
$\frac{g_1}{f_1}=\frac{G_A}{G_V}$ at $Q^2$= 0. The
$\chi$CQM$_{{\rm config}}$ is able to give a fairly good account
for most of the decays where the experimental data is available
for example, in the case of ${\frac{G_A}{G_V}}^{\Sigma^-
\rightarrow n}$, ${\frac{G_A}{G_V}}^{\Xi^- \rightarrow \Lambda}$,
${\frac{G_A}{G_V}}^{\Lambda \rightarrow p}$ and
${\frac{G_A}{G_V}}^{\Xi^0 \rightarrow \Sigma^+}$. The SU(3)
symmetry breaking as well as configuration mixing effects have
been incorporated in the calculations and it clearly leads to a
better agreement with data. Therefore, both are simultaneously
important to obtain the desired agreement. A refinement in the
case of the measurements with the assumption of SU(3) symmetry
breaking would have important implications for the basic tenets of
$\chi$CQM as well as SU(3) symmetry breaking.

In conclusion, we would like to state that at the leading order
constituent quarks and the weakly interacting Goldstone bosons
constitute the appropriate degrees of freedom in the
nonperturbative regime of QCD  in the $\chi$CQM$_{{\rm config}}$
and SU(3) symmetry breaking is the key in understanding the spin
content and the weak axial-vector form factors of the
semi-leptonic octet baryon decays.

 \vskip .2cm
 {\bf ACKNOWLEDGMENTS}\\
H.D. would like to thank CCSTDS, Government of India and the
organizers of SPIN2008, University of Virginia, for financial
support.

%\pagebreak

\begin{table}
\begin{center}
\begin{tabular}{lcccccc} \hline
Decay & $ f_1$  & $f_2$ & $f_3$ & $g_1$ & $g_2$ & $g_3 $\\ \hline
$n\rightarrow p e^{-} \bar{\nu}$ & 1.00 & 2.612 & 0.003& 1.270
&$-$0.004& $-$232.9\\ $ \Sigma^- \rightarrow \Sigma^0 e^{-}
\bar{\nu} $ & 1.414 & 1.033 & 0.005 & 0.676 & $-$0.010 & $-$201.3
\\ $ \Sigma^-\rightarrow \Lambda e^{-} \bar{\nu}$&0 & 2.265 &
0.080 & 0.646 & $-$0.152 & $-$271.4\\ $  \Sigma ^+ \rightarrow
\Lambda e^{-} \bar{\nu}$ &0 & 2.257 & 0.072 & 0.646 & $-$0.136 &
$-$245.9\\ $ \Xi^- \rightarrow \Xi^0 e^{-} \bar{\nu} $ & $-$1.00 &
2.253 & 0.003 & 0.314 & $-$0.007 & 113.8\\ \hline
$\Sigma^-\rightarrow n e^{-} \bar{\nu}$ & $-$1.0 & 1.813 & 0.616 &
0.314 & 0.017 & $-$9.2\\ $\Xi^-\rightarrow \Sigma^0 e^{-}
\bar{\nu}$ & 0.707 & 2.029 & $-$0.291 & 0.898 & 0.310 & $-$29.1\\
$\Xi^-\rightarrow \Lambda e^{-} \bar{\nu}$ & 1.225 & $-$0.450 &
$-$0.658 & 0.262 &0.047 & $-$8.9 \\ $\Lambda \rightarrow p e^{-}
\bar{\nu}$& $-$1.225 & $-$1.037 & 0.415 & $-$0.909 & $-$0.170 &
20.7\\ $ \Xi^0 \rightarrow \Sigma^+ e^{-} \bar{\nu}$& 1.0 & 2.854
& $-$0.414 & 1.27 & 0.446 & $-$40.7\\ \hline
 \end{tabular}\\
\caption{Weak vector and axial-vector form factors for the
semi-leptonic octet baryons decays in the $\chi$CQM$_{{\rm
config}}$.} \label{fsandgs} \end{center}
\end{table}

\begin{table}
\begin{center}
\begin{tabular}{lcc  cc} \hline
Decay & Data& NQM      &$\chi$CQM$_{{\rm config}}$&
$\chi$CQM$_{{\rm config}}$ with \\ &\cite{PDG} &  & with SU(3) &
SU(3) symmetry \\
 & &
 & symmetry &  breaking \\      \hline
$\frac{G_A}{G_V}^{n\rightarrow p}$ & 1.2695 $\pm$ 0.0029  &1.67 &
0.95& 1.27\\ $\frac{G_A}{G_V}^{\Sigma^- \rightarrow \Sigma^0} $ &
-- & 0.67 & 0.39&0.48\\ $\frac{G_A}{G_V}^{\Sigma^-\rightarrow
\Lambda}$  & ${\frac{f_1}{g_1}=0.01\pm 0.1 }$ &0.82  & 0.45&
0.65\\
 $\frac{G_A}{G_V}^{\Sigma^+ \rightarrow \Lambda}$ &--&0.82   &0.45 & 0.65\\
 $\frac{G_A}{G_V}^{\Xi^- \rightarrow \Xi^0} $ &--& $-$0.33&$-$0.16 &$-$0.31\\\hline

$\frac{G_A}{G_V}^{ \Sigma^-\rightarrow n}$ & $-$0.340$\pm$ 0.017
&$-$0.33&$-$0.16 &$-$0.31\\ $\frac{G_A}{G_V}^{\Xi^-\rightarrow
\Sigma^0}$ &-- & 1.67& 0.95&1.27\\
$\frac{G_A}{G_V}^{\Xi^-\rightarrow \Lambda}$ & 0.25 $\pm$0.05
&0.33  &0.21 &0.21\\
 $\frac{G_A}{G_V}^{\Lambda \rightarrow p}$ &0.718$\pm$ 0.015 & 1.00&0.58& 0.74\\
 $\frac{G_A}{G_V}^{\Xi^0 \rightarrow \Sigma^+ }$&1.21$\pm$0.05 &1.67 &0.95&1.27\\ \hline

 \end{tabular}
 \caption{Predictions for the the axial-vector form factors $G_A/G_V$ in $\chi$CQM$_{{\rm
config}}$.} \label{g1f1} \end{center}
\end{table}

\end{document}